\documentclass[11pt, a4paper]{article}
\usepackage[margin=1in]{geometry}

\usepackage{array}    
\usepackage[whole]{bxcjkjatype}
\usepackage{booktabs} % \toprule, \midrule, \bottomrule
\usepackage{tabularx} % tabularx 
\usepackage{multirow} % \multirow 
\usepackage{array}    % \newcolumntype
\usepackage{url}
\usepackage{graphicx} % for preprint
\usepackage{authblk}  % for preprint

\newcolumntype{Y}{>{\centering\arraybackslash}X}

\begin{document}

%%%% Article title to be placed here
\title{Agency at the Interface: Distinguishing Teleological from Structural Self-Organization via Internal Coarse-Graining and Downward Causation}

%\title{Agency at the Interface: A Duality Framework of Coarse-Graining and Downward Causation}

% for preprint
%\author{%%%% Author details
%Kazuya Horibe$^{1}$, Keisuke Suzuki$^{2}$ }

%%%%%%%%% Insert author address here
%for preprint
%\address{$^{1}$RIKEN Center for Brain Science, Wako, Japan\\
% $^{2}$Center for Human Nature, Artificial Intelligence, and Neuroscience (CHAIN), Hokkaido University, Sapporo, Japan}

% for preprint
\author[1]{Kazuya Horibe}
\author[2]{Keisuke Suzuki}

\affil[1]{RIKEN Center for Brain Science, Wako, Japan}
\affil[2]{Center for Human Nature, Artificial Intelligence, and Neuroscience (CHAIN), Hokkaido University, Sapporo, Japan}

% for preprint
\date{}
%%%% Subject entries to be placed here %%%%
% for preprint
%\subject{Artificial Life, Artificial Intelligence}

%%%% Insert corresponding author and its email address}
% for preprint
%\corres{Kazuya Horibe\\
%\email{kazuya.horibe@riken.jp}}

%%%% Abstract text to be placed here %%%%%%%%%%%%
% for preprint
\maketitle
\begin{abstract}
Agency is widely characterized as the capacity of a system to regulate its internal states toward self-generated goals, yet characterizing the functional basis of this autonomy requires a distinction between the system's own organization and an observer's interpretation. In this paper, we ground this capacity functionally in the generative process of intrinsic dynamics, characterized as a self-referential loop generated by internal coarse-graining and downward causation. This process allows the system to autonomously compress microscopic states into macroscopic variables that subsequently constrain microscopic temporal evolution. By distinguishing the intrinsic dynamics of the system from the external coarse-graining of an observer's interpretation, we define agency through the dynamics of the predictive gap. This gap constitutes the internal divergence between the system's anticipation and its realization, as well as the limited reducibility of the system's generated constraints within the observer's interpretive model. This framework outlines a spectrum of agency that distinguishes biological systems, whose teleological constraints are generated through their own regulatory dynamics, from artificial systems, whose organization is designed to satisfy externally specified objectives. Finally, we extend this interface to the social domain, proposing that these generative divergences underpin participatory sense-making and emergent coordination.

% This duality framework could offer a spectrum for classifying agency in natural and artificial systems and reframe social interaction as reciprocal modeling and emergent entanglement.

\end{abstract}
%%%%%%%%%%%%%%%%%%%%%%%%%%%

%%%% Keyword entries to be placed here %%%%
% for preprint
%\keywords{Agency, Internal Coarse-Graining, Downward Causation, Good Regulator Theorem, Teleology}

% for preprint
%\rsbreak

%%%%%%%%%% Insert the texts which can accomdate on firstpage in the tag ``fmtext`` %%%%%

%%%%%%%%%%%%%%% End of first page %%%%%%%%%%%%%%%%%%%%%
% for preprint
%\maketitle
\section{Introduction} \label{sec: introduction}

Agency, the capacity for goal-directed behavior, is a fundamental concept to understanding living and intelligent systems \cite{maturana1970biology,maturana1980autopoiesis,di2014enactive,barandiaran2009defining, tomasello2022evolution,ginsburg2019evolution,feinberg2017ancient}. Unlike passive systems such as a rolling stone governed by physical laws, agents regulate their internal states to pursue self-generated goals \cite{cannon1939wisdom,ashby1952design}, highlighting the core property of autonomous intrinsic dynamics \cite{varela1979principles,moreno2015biological}. Recently, artificial systems, including Large Language Models (LLMs) capable of fluent dialogue  \cite{achiam2023gpt,guo2025deepseek,comanici2025gemini,yang2025qwen3} and robots with smooth and lifelike locomotion \cite{unitree2024g1,unitree2023go2,boston2020spot}, have begun to blur this boundary. As the mechanisms of these artificial systems become increasingly difficult for humans to comprehend fully, their behavior can be interpreted as exhibiting autonomy similar to that of living systems, suggesting the need for a hybrid conception of agency that spans both \cite{baltieri2023hybrid}.

Formalizing agency has been a pursuit across multiple disciplines \cite{baltieri2025_mathematical}, encompassing dynamical systems approaches \cite{beer1995dynamical}, information-theoretic approaches \cite{krakauer2020information, kenton2023discovering}, and causal structure approaches like Integrated Information Theory \cite{aguilera2018agency, desmond2022integrated, albantakis2021quantifying,albantakis2023integrated}. Prominent among these are the Free Energy Principle (FEP), which characterizes agents as minimizing variational free energy \cite{friston2006free,friston2010free,friston2019free}, and Good Regulator Theory (GRT), which posits that every successful regulator must constitute a model of its system \cite{conant1970every,ashby1952design}, particularly because they establish a formal correspondence between agency and epistemology, asserting that the functional capacity of regulation is mathematically equivalent to possessing a predictive model of the environment.

%particularly because both principles link world-predictive internal models to a system's own regulatory capacities, offering a distinctive way of understanding such systems as agents.

Despite these contributions, the generality of these frameworks introduces a persistent ambiguity between an external observer's description and a system's intrinsic dynamics \cite{ramstead2020tale}. For instance, under the FEP, even simple feedback systems can be mathematically interpreted as performing Bayesian inference based on a generative model ascribed by the observer rather than implemented by the system itself \cite{mcgregor2017bayesian,baltieri2020predictions}. Similarly, counterexamples to GRT, such as Braitenberg vehicles \cite{braitenberg1986vehicles,beer2003dynamics}, make this ambiguity explicit: their seemingly goal-directed regulation arises from an observer's interpretation rather than from any internal model implemented by the system's own intrinsic dynamics. These ambiguities indicate that while FEP and GRT serve as powerful interpretive tools, they currently lack the formalization necessary to establish a correspondence between the intentional stance (later referred to as the ``as-if'' stance)\cite{dennett1989intentional} and the intrinsic dynamics of systems.

To resolve this ambiguity and provide the missing formalization, recent frameworks utilizing interpretation maps and Bayesian filtering have mathematically grounded the relationship between intrinsic dynamics and the observer's inferential description \cite{virgo2021interpreting,biehl2022interpreting,virgo2025good,baltieri2025bayesian}. By establishing a formal correspondence between the mechanics of regulation and the logic of modeling, these works demonstrate that the functional capacity to regulate is mathematically equivalent to possessing a model of its own, even in systems lacking explicit representations like Braitenberg vehicles. This perspective helps clarify the relationship between the ``as-if'' stance of the model with the intrinsic dynamics of systems. However, these frameworks remain primarily descriptive; they take the existence of a successful regulator as a premise. Consequently, the generative mechanisms, i.e., how such regulating systems arise, remain outside the theoretical scope.

To complement this epistemic interpretative framing, the functional characterization of autonomous agents as far-from-equilibrium systems offers a generative perspective, emphasizing active self-maintenance via intrinsic dynamics \cite{ruiz2004basic,thompson2009making,froese2009enactive,maes1990designing}. A potentially key pathway toward such self-maintenance is hypothesized to be the compression of system-specific macroscopic variables through internal coarse-graining, which functions as a form of downward causation \cite{flack2017coarse}. In this hypothesis, the internalization of intrinsic variables (internal coarse-graining) and the subsequent causal constraints (downward causation) constitute a self-referential loop, potentially serving as the functional source of the system's capacity to regulate \cite{juarrero2000dynamics}.

Based on this perspective, we propose a duality framework that grounds agency in the distinction between observer and system, explicitly addressing the generative process of intrinsic dynamics.
%Based on this perspective, we propose a duality framework of agency designed to resolve the ambiguity between observer and system by explicitly addressing the generative process of intrinsic dynamics.
First, we incorporate internal coarse-graining and downward causation into the definition of agency, providing a functional grounding for the intrinsic dynamics that underpin self-regulation capacity (Section \ref{sec: system}). Second, we make a clear distinction between system and observer for distinguishing intrinsic dynamics from external descriptions. Accordingly, we characterize agency by the gap between these intrinsic dynamics and the observer's ``as-if'' model (Section \ref{sec: interface and observer}). We then apply this framework to various systems to explore an agency assessment that reflects a specific autonomous structure, which we frame as a spectrum ranging from structural to teleological self-organization, distinguishing mere pattern formation from internally guided self-maintenance (Section \ref{sec: spectrum}). Finally, in the Discussion, we extend our framework to the social domain, discussing how the maintenance of the internal-external gap might distinguish social entanglement from mere mechanical synchronization (Section \ref{sec: discussion}).

\begin{figure}
    \centering
    \includegraphics[width=1.0\linewidth]{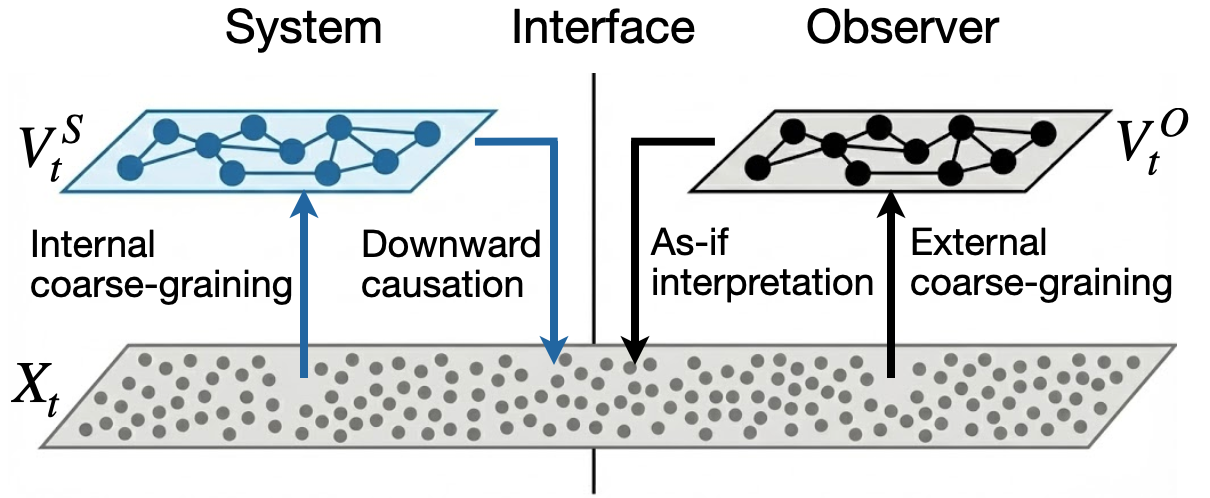}
    \caption{\textbf{A Duality Framework of Agency.} The diagram shows the structural parallel between the system's intrinsic dynamics (left) and the observer's interpretive model (right). The System (blue) operates through a self-referential loop where internal coarse-graining maps microscopic states $X_t$ to intrinsic macroscopic variables $V_t^S$, which then influence the micro-level via downward causation. The Observer (black) operates through an epistemic loop where external coarse-graining maps the same microscopic states to observational variables $V_t^O$, closing the loop via interpolation. The central vertical line demarcates the interface between the system's intrinsic dynamics and the observer's epistemic description.}
    \label{fig: conceptual framework}
\end{figure}

\section{A Generative Process of Intrinsic Dynamics} \label{sec: system}
To establish the functional basis for the generative process of intrinsic dynamics, we characterize the system through the self-referential loop between its microscopic dynamics $X_t$ and intrinsically generated macroscopic variables $V_t^S$ (Figure \ref{fig: conceptual framework}, System). Microscopic variables $X_t$ typically correspond to elementary components governed by fundamental physical laws or local interaction rules, such as individual molecules in biochemical reactions. In contrast, macroscopic variables $V_t^S$ are intrinsically generated quantities that screen off specific microscopic details to obey their own macroscopic constraints. These variables may manifest as ``fragments'' serving as functional units of interaction \cite{feret2009internal}, slow order parameters that enslave fast dynamical modes \cite{haken1987information}, or macroscopic structures acting as boundary conditions \cite{kondo1995reaction,levin2017endogenous}.

This self-referential loop is mediated by two mechanisms; the bottom-up process of internal coarse-graining, where the system autonomously compresses microscopic details into macroscopic states (Figure \ref{fig: conceptual framework}, Internal coarse-graining), and the top-down process of downward causation, whereby these intrinsically generated variables contract the effective structure, acting as operators to constrain the microscopic dynamics (Figure \ref{fig: conceptual framework}, Downward causation).

\subsection{Internal Coarse-Graining as the Generative Basis of Intrinsic Structure
} \label{sec: internal coarse-graining}
%Internal Coarse-Graining: The Emergence of Intrinsic Macroscopic Variables

Internal coarse-graining denotes the process by which a system autonomously generates its own macroscopic variables $V_{t}^{S}$ through its organization \cite{flack2017coarse} (Figure \ref{fig: conceptual framework}, Internal coarse-graining). This concept differs fundamentally from conventional coarse-graining employed in statistical physics \cite{anderson1972more,kadanoff1966scaling,jaynes1957information}. While the latter functions as an analytical tool for observers to abstract away microscopic complexity, internal coarse-graining represents a dynamical feature generated by the system itself, independent of external description.

In molecular biology, cellular signaling networks show internal coarse-graining by operating effectively despite the astronomical number of potential molecular species, without requiring their explicit enumeration. Rule-based models of biochemical signaling (e.g., in the Kappa language) formalize this principle underlying internal coarse-graining, showing that the system's own interaction rules determine which distinctions are dynamically effective \cite{feret2009internal}. The resulting coarse-grained variables, termed ``fragments,'' capture the finest level of resolution that the system's dynamics can effectively distinguish. Rather than being abstract summaries defined by an observer, these fragments emerge as the specific macroscopic variables $V_t^S$ that the system itself ''sees'' and maintains amidst microscopic noise. A related phenomenon is the slaving principle in synergetics \cite{haken1987information}, where systems near critical instabilities self-organize into a few slow, macroscopic order parameters and many fast, slaved modes. The fast modes lose independent dynamics and become determined by the slow variables, demonstrating an intrinsic time-scale separation that compresses microscopic information. 

Internal coarse-graining is also present in artificial systems when the architecture forces the system to identify and stabilize its own macroscopic variables\footnote{Because such objectives are assigned rather than self-generated, these cases fall under structural self-organization, in contrast to the teleological organization characteristic of biological systems.}. For instance, a Variational Autoencoder (VAE) is driven by the objective of minimizing reconstruction error to autonomously compress high-dimensional inputs $X_t$ into a compact latent representation\cite{kingma2013auto}. Here, the latent vector functions as the macroscopic variable $V_t^S$ generated by the system; it is not merely a data compression but an internal identification of the essential low-dimensional structure required to satisfy the system's objective. Similarly, feed-forward (FF) layers in Transformer networks act as key-value memories that reorganize semantically relevant information \cite{geva2021transformer}, effectively filtering microscopic token details into semantically consolidated macroscopic states essential for the network's operation.

% From the observer's perspective, the Information Bottleneck (IB) principle 
To epistemically evaluate, the efficacy of this intrinsic compression, the Information Bottleneck (IB) principle \cite{tishby2000information,shwartz2017opening} provides an analytical lens. This principle formalizes the trade-off between discarding irrelevant microscopic details and preserving functionally relevant information. Within this framework, an optimal macroscopic variable $V_{t}^{S}$ is defined as one that balances two competing objectives: maximizing the compression of the detailed micro-state while maximizing the information preserved regarding a specific target consequence. In our context, IB allows the observer to verify whether the system has successfully extracted efficient macroscopic representations relevant to its function.

While metrics like the Information Bottleneck (IB) provide an epistemic indication of compression, one should not confuse such indicators with the intrinsic macroscopic variables themselves, which are realized through biochemical or computational mechanisms. Although observers must ``discover'' these variables within the architecture, this identification is distinct from merely characterizing statistical properties. Rather than positing distinct ontological substances, we identify these variables as functional entities intrinsic to the mechanism, which operate as context-dependent regulators governing the system's dynamics. Crucially, this generation constitutes only the first half of the self-referential loop. The question of how these generated variables subsequently constrain the system constitutes the second half of the loop: downward causation.

% in biological systems is realized through biochemical mechanisms that constrain chemical reaction dynamics, while in artificial systems, it is implemented through computational mechanisms that organize latent spaces in neural networks. 
%Internal coarse-graining, which is realized through biochemical or computational mechanisms should be not be confused with the information theoritical measures like IB. Although high performance on IB objectives provides an epistemic indication that useful compression has been achieved, such performance reflects only the observer's evaluation. Internal coarse-graining must function as an observer-independent process, in which the system's own dynamics generate the macroscopic variables $V_t^S$. Crucially, this generation constitutes only the first half of the self-referential loop. The question of how these generated variables subsequently constrain the system constitutes the second half of the loop: downward causation.

\subsection{Downward Causation as Structural and Dynamical Constraints} \label{sec: Downward Causation}

%Internal coarse-graining generates macroscopic variables from microscopic dynamics. To understand how these macroscopic variables subsequently influence the system, we draw upon a distinction developed in the philosophy of emergence between synchronic and diachronic forms of downward causation \cite{kim1999making}. This distinction corresponds to what has been characterized as the difference between vertical determination, which concerns relationships between micro and macro levels, and horizontal causation, which concerns the temporal flow from past to present to future \cite{kim2005physicalism}.
%Once generated through internal coarse-graining, the macroscopic variables operate as context-dependent constraints \cite{juarrero2000dynamics, juarrero2023context}, serving as functional regulators that limit the degrees of freedom of the lower-level microscopic dynamics. This process, known as downward causation, constitutes the second phase of the self-referential loop and is widely observed in biological systems (Figure \ref{fig: conceptual framework}, Downward causation).

Downward causation occurs when the macroscopic variables $V_t^S$ generated through internal coarse-graining begin to act as context-dependent constraints on the microscopic dynamics $X_t$, rather than remaining passive descriptions. This process constitutes the second phase of the self-referential loop: downward causation\cite{kim1998mind}. To characterize how these constraints operate, we draw upon the distinction in the philosophy of emergence between synchronic and diachronic forms of determination \cite{sartenaer2015synchronic}. This distinction provides a theoretical basis for categorizing downward causation into two complementary modes: structural constraints, which specify the system's instantaneous organization, and dynamical constraints, which govern its temporally extended evolution. It is crucial to clarify that we employ `downward causation' strictly in its organizational or `weak' sense \cite{emmeche2000levels}: we do not posit new physical forces that violate microscopic laws, but rather constraints that restrict the degrees of freedom of microscopic components \cite{juarrero2000dynamics, juarrero2023context}, the system's dynamics toward coherent patterns and coordinated microscopic interactions that reinforce the macroscopic organization.

% Internal coarse-graining produces macroscopic variables that, through complementary forms of downward causation \cite{kim1998mind,sartenaer2015synchronic}, function as context-dependent constraints. These constraints structure the system's instantaneous configuration \cite{juarrero2000dynamics, juarrero2023context} (Figure \ref{fig: time evolution}). It is crucial to clarify that we employ 'downward causation' strictly in its organizational or 'weak' sense \cite{emmeche2000levels}: we do not posit new physical forces that violate microscopic laws, but rather constraints that restrict the degrees of freedom of microscopic components, thereby channeling the system's dynamics toward specific functional outcomes.

% This structural constraint ensures that local dynamics remain compatible with the macroscopic boundary conditions, yielding local dynamics that remain organized by the macroscopic boundary conditions rather than reducing to a set of unconstrained microscopic state transitions. This mechanism constitutes low-level downward causation, in which macroscopic variables serve as constraining conditions that select among physically possible trajectories without violating physical laws \cite{emmeche2000levels}. Rather than acting as a new physical force, these constraints operate by reducing the degrees of freedom of the microscopic components, channeling their dynamics toward specific functional outcomes. 

Structurally, macroscopic variables $V_t^S$ serve as a boundary condition that restricts the range of configurations available to the microscopic variables $X_t$. This ensures that local dynamics do not evolve as an aggregate of isolated events but remain subject to context-dependent constraints imposed by the macroscopic organization. Biological morphogenesis shows this spatial constraint, where global patterns like bioelectrical gradients \cite{Turing1952chemical, kondo1995reaction, levin2017endogenous} or tissue curvature \cite{taniguchi2013phase, horibe2019curved, nishide2022pattern} bias local gene expression toward possibilities compatible with global geometry. Similarly, in artificial systems like the Game of Life, system-generated structures (e.g., gliders) constrain the system not by altering local rules, but by determining the sequence of local neighborhoods encountered by individual cells \cite{beer2004autopoiesis, beer2014cognitive}. Analogous constraints are found in self-replicating loop networks where global patterns stabilize replication \cite{suzuki2006spatial}, and in protocell models where membranes spatially restrict internal reactions \cite{varela1974autopoiesis, suzuki2009shapes}. In these cases, the macro-level organization does not violate micro-rules but shapes the context in which they operate.

% Biological morphogenesis shows this spatial constraint; global patterns such as bioelectrical gradients \cite{Turing1952chemical,kondo1995reaction, levin2017endogenous} and tissue curvature serve as macroscopic boundary conditions \cite{taniguchi2013phase} that bias gene expression toward possibilities compatible with global geometry. Similarly, as the tissue deforms, the global configuration reorganizes and further constrains local processes \cite{taniguchi2013phase, horibe2019curved, nishide2022pattern}. Artificial systems exhibit similar constraints; for instance, in the Game of Life, the persistence of a glider arises because the global configuration determines the sequence of local neighborhoods encountered by individual cells\cite{beer2004autopoiesis, beer2014cognitive}, and in self-replicating loop networks, emergent macroscopic structures constrain the stability of replication of loops \cite{suzuki2006spatial}.

Dynamically, the system's macroscopic organization guides its temporal evolution. Unlike passive systems determined exclusively by constituent microscopic dynamics, an intrinsic dynamics $F_{S}$ acts as a global constraint that biases transition probabilities \cite{kim1999making}. This biasing allows for anticipation $V_{t'}^{S_{ant}}$, utilizing context-sensitive constraints like rhythm and timing \cite{juarrero2015does}. This becomes operational in bacterial genetic networks, where \textit{E. coli} uses temperature elevation to suppress respiration genes in anticipation of oxygen depletion \cite{tagkopoulos2008predictive}. This regulatory loop indicates that the transition rule operates as an organizational constraint whose form is not fixed by underlying physical laws. The fact that bacteria can evolve to weaken this association when environmental correlations reverse shows that the constraint is an intrinsic dynamics of the system's organization rather than a pattern imposed by passive dynamics alone. Formally, this targeted future state $V_{t'}^{S_{ant}} = F_{S}(V_t^{S})$ represents a teleological reference point that imposes a selective pressure on the actualized microscopic temporal evolution.

From an analytical perspective, the operation of this self-referential loop can be verified using information-theoretic tools that track how macroscopic organization shapes microscopic temporal evolution.
%From the observer's perspective, the system's intrinsic dynamics can be quantified using information-theoretic tools that track how macroscopic organization shapes microscopic evolution. 
First, the validity of internal coarse-graining is captured by Effective Information \cite{hoel2013quantifying}, which identifies when a macroscopic variable $V_t^S$ displays a more coherent causal organization than the underlying microscopic state $X_t$. Second, the structural efficacy of these variables, specifically how they act as boundary conditions constraining the microscopic phase space, is formalized through the Downward Causation index \cite{rosas2020reconciling}. Third, the dynamical autonomy of the system, defined as whether the temporal evolution from $V_t^S$ to $V_{t'}^S$ follows a macroscopic rule irreducible to micro-level details, is assessed using metrics such as Dynamical Independence \cite{barnett2023dynamical} and Computational Closure \cite{rosas2024software}. Collectively, these measures delineate when the interplay of structural constraints and temporal autonomy forms a coherent self-referential loop. 
While they offer a methodology to characterize the system's organization from an ``as-if'' stance, unlike the biochemical or computational mechanisms identified above as functional entities, these metrics operate on random variables as probabilistic descriptions of relationships between variables.
% From the observer's perspective, the system's intrinsic dynamics can be quantified using a set of quantitative tools that track how macroscopic organization shapes microscopic evolution. Internal coarse-graining provides the first step; the high-dimensional microscopic state $X_t$ is summarized into an intrinsic macroscopic variable $V_t^S$, which often displays a more coherent causal organization than the underlying microstates, as captured by Effective Information \cite{hoel2013quantifying}. Once defined, $V_t^S$ can be used to evaluate Synchronic Downward Causation, which quantifies how macroscopic boundary conditions constrain microscopic phase space, formalized through the Downward Causation index $\mathcal{D}$ \cite{rosas2020reconciling}. Diachronic Downward Causation concerns whether the evolution from $V_t^S$ to $V_{t'}^S$ follows an autonomous macroscopic rule rather than being reducible to micro-level details, assessed using metrics such as Dynamical Independence \cite{barnett2023dynamical} and Computational Closure \cite{rosas2024software}. Collectively, these observer-oriented measures delineate when coarse-graining and downward causation form a coherent self-referential loop. While they do not reveal intrinsic mechanisms, they offer a methodology to characterize the system's organization from an ``as-if'' stance.

Taken together, the generative processes of intrinsic dynamics $F_{S}$ described above operate in concert: internal coarse-graining, structural and dynamical downward causation jointly establish a closed self-referential loop that links the microscopic variables $X_t$ with the macroscopic organization $V_t^{S}$. This loop reflects a structured interaction in which macroscopic variables are generated, exert constraints, and stabilize their own temporal evolution.

\subsection{Summary}
In summary, the generative process of intrinsic dynamics is defined by a self-referential loop where internal coarse-graining and downward causation reciprocally determine one another. 
By autonomously compressing microscopic complexity, the system generates macroscopic variables ($V_{t}^{S}$) that act as active constraints, shaping instantaneous configurations and guiding temporal evolution. We see this in bacteria that utilize temperature as a macroscopic proxy to anticipate oxygen depletion, demonstrating how intrinsic constraints enable proactive regulation. Seen in this light, the framework developed here provides a generative interpretation of the intuition formalized in Good Regulator Theory. What appears, from the observer's perspective, as the system having an effective regulatory model can instead be understood as the outcome of a self-referential loop in which the system constructs macroscopic variables and uses them to shape its own dynamics. Thus, the ``good regulator'' need not be treated as a pre-given property but may instead be interpreted as an organizational pattern that could arise from intrinsic dynamics.

%Through internal coarse-graining, the system autonomously compresses microscopic complexity into intrinsic macroscopic variables ($V_{t}^{S}$), distinct from observer-defined abstractions. These variables subsequently function as constraints on the system itself: they serve as boundary conditions restricting instantaneous microscopic configurations or define the transition rules that guide the system's temporal evolution. This mechanism is exemplified by bacteria, which have evolutionarily acquired temperature as a macroscopic variable to proactively regulate gene networks in anticipation of oxygen depletion. This suggests that the "good regulator" described in Good Regulator Theory was autonomously acquired by the bacterium itself through the dual processes of internal coarse-graining and downward causation, thereby demonstrating a potential generative mechanism for the emergence of such regulatory capacities.

\begin{figure}
    \centering
    \includegraphics[width=1.0\linewidth]{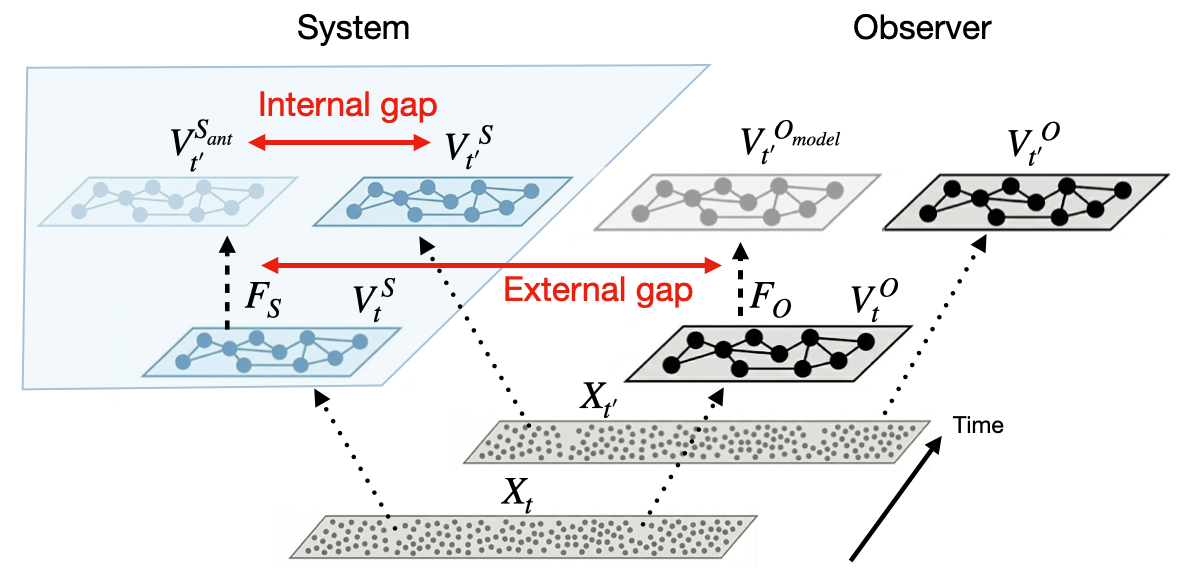}
    \caption{\textbf{The Duality of Gaps arising from Divergent Coarse-Graining.} The diagram contrasts the temporal evolution ($t \rightarrow t'$) of the system's intrinsic dynamics (left panel, blue) with the observer's interpretive model (right panel, black). The red double-headed arrows denote the two structural divergences: the Internal Gap between the system's anticipation ($V_{t'}^{S_{ant}}$) and the realized state ($V_{t'}^{S}$), and the External Gap between the intrinsic dynamics ($F_{S}$) and the observer's model ($F{O}$). Notation: $X$ and $V$ denote microscopic and macroscopic variables, respectively. Solid arrows: Realized state transitions. Dotted arrows: Coarse-graining mappings. Dashed arrows: Anticipated or predicted trajectories.}
    \label{fig: time evolution}
\end{figure}

\section{The Observer-System Interface and the Attribution of Agency} \label{sec: interface and observer}
Having outlined how intrinsic dynamics give rise to regulatory structure, we now turn to how such systems are interpreted from the outside. To analyze systems that potentially instantiate a self-referential loop by ``as-if'' stance, an observer applies epistemic coarse-graining to the microscopic dynamics $X_t$ of the system. This process generates a distinct set of macroscopic variables $V_t^O$ and constructs an ``as-if'' model $F_{O}$ to describe the behavior (Figure \ref{fig: conceptual framework}, Observer). This interpolation by the observer necessitates accounting for two gaps. The first is the internal gap between the macroscopic states $V_t^S$ of the system and its own anticipation $V_{t}^{S_{ant}}$ (Figure \ref{fig: time evolution}, Internal Gap). The second is the external gap between the intrinsic dynamics of the system $F_{S}$ and the model constructed by the observer $F_{O}$ (Figure \ref{fig: time evolution}, External Gap).

\subsection{The Internal Gap as the Divergence between Anticipation and Realization} \label{sec: Gap between micro and macro}
%ksk
% Within the self-referential loop, the Internal Gap characterizes the divergence between the system's intrinsic anticipation and its realized states. Given the macroscopic state $V_{t}^{S}$ produced by internal coarse-graining, the intrinsic dynamics $F_{S}$ specify an anticipated successor state $V_{t'}^{S_{ant}}$. However, the microscopic variables $X_t$ update under noise and environmental perturbations, producing a realized state $X_{t'}$ that yields the actual macroscopic outcome $V_{t'}^{S}$ when coarse-grained. The Internal Gap is formally defined as the divergence $D(V_{t'}^{S_{ant}} \| V_{t'}^S)$ between these two outcomes. Biological systems maintain a non-zero gap that functions as a critical error signal, providing the informational basis to adjust microscopic processes. Ultimately, the system's persistence relies on regulating this deviation to remain within bounds compatible with its self-organization.
The Internal Gap marks the interface between the system's projected teleology and its physical realization (Figure \ref{fig: time evolution}). Formally, this gap quantifies the divergence between the state anticipated by the system's intrinsic dynamics $V_{t'}^{S_{ant}}$ and the state actually realized by its microscopic substrate $V_{t'}^S$. This divergence arises because the intrinsic dynamics $F_{S}$ act as a regulatory operator that delineates the state transition tendencies encoded in the system's organization, while the microscopic substrate $X_t$ updates under stochastic noise. A bacterium's chemotactic control, discussed earlier, exemplifies this relation without requiring further elaboration here. Unlike passive systems determined exclusively by constituent micro-dynamics, biological systems maintain this gap as a functional space for exercising agency. Here, the divergence serves as a feedback signal: when the realized state deviates from the anticipation, it triggers regulatory corrections. Ultimately, the system's persistence is defined by its capacity to manage this divergence, keeping it within bounds compatible with its self-organization.

This capacity to regulate the Internal Gap is shown in cases where organisms actively move away from harmful chemical concentrations, an adaptive response known as negative chemotaxis. Under typical conditions, amino acids and sugars function as attractants that the intrinsic dynamics $F_{S}$ treat as signals of favorable conditions. When these same compounds reach concentrations that compromise osmotic balance or pH homeostasis, the realized macroscopic state $V_{t'}^{S}$ no longer aligns with the anticipated state $V_{t'}^{S_{ant}}$. In such cases, the bacterium adjusts its microscopic processes so that its movement reverses away from the source \cite{tso1974negative, mao2003sensitive}. A similar pattern is observed in the acute avoidance of Reactive Oxygen Species (ROS), where the intrinsic dynamics generate an anticipation of cellular damage that leads the realized state to shift toward escape behaviors \cite{benov1996escherichia}. In both instances, the Internal Gap functions not merely as a gradient-tracking error but as a higher-level regulatory interface, where the macroscopic imperative for survival reorganizes the microscopic responses through the constraints revealed by the Internal Gap.

% This capacity to manage the gap is strikingly illustrated by negative chemotaxis, where the system acts to preserve its integrity even against immediate nutrient incentives \cite{tso1974negative}. Typically, amino acids and sugars serve as attractants that the system anticipates as beneficial. However, when these concentrations reach toxic levels that threaten osmotic balance or pH homeostasis, the bacterium autonomously inverts its response to actively flee the source \cite{mao2003sensitive}. Similarly, the acute avoidance of Reactive Oxygen Species (ROS) demonstrates a survival-driven hierarchy: the intrinsic dynamics ($F_S$) generate an anticipation of lethal damage ($V_{t'}^{S_{ant}}$) that overrides the local drive for nutrients \cite{benov1996escherichia}. In these instances, the Internal Gap functions not merely as a gradient-tracking error but as a high-order decision interface, where the macroscopic imperative for survival constrains and reconfigures microscopic reflex arcs.

Crucially, the persistence of the Internal Gap, even when regulation fails, distinguishes a pathological agent from a passive systems. Regulatory failure can be classified into two general forms. These modes correspond to breakdowns in the self-referential loop's two essential capacities: implementing appropriate microscopic adjustments (loss of fidelity) and revising anticipatory structures in response to persistent mismatches (loss of plasticity). The first is loss of fidelity, where the system generates appropriate anticipations but cannot implement the necessary microscopic adjustments. In somatic conditions like Chronic Fatigue Syndrome (CFS), inflammatory processes disrupt the mechanisms needed to meet anticipated energetic demands, causing a persistent divergence between anticipation and realization \cite{santamaria2025allostatic,hogan2020neural}. The second is loss of plasticity, where the system fails to update its anticipatory structure despite repeated mismatches. In psychiatric disorders such as anxiety, the system assigns excessive precision to threat-related predictions, driving behavior through fixed expectations rather than actual conditions \cite{paulus2019active,barrett2016active}. Taken together, these forms of regulatory failure indicate that even when a system cannot effectively use the Internal Gap to guide its behavior, the continued presence of this gap marks a structured, self-referential organization that remains distinct from purely passive dynamics.

In those two modes of failure, we propose that the formal differentiation between $V_{t'}^{S_{ant}}$ and $V_{t'}^{S}$ represents the functional interface where the system's self-referential loop attempts to regulate its own organization. While the Good Regulator Theorem implies that effective regulation necessitates a requisite internal model (suggesting that failure reflects a loss of this isomorphism), the Internal Gap framework suggests a critical refinement: the persistent maladaptation described above—whether due to lost fidelity or plasticity—can be interpreted not as the mere absence of an internal model, but as the activity of intrinsic dynamics $F_{S}$ attempting regulation that has failed. Within this view, the continued existence of the gap implies that the system retains its teleological orientation, distinguishing regulatory dysfunction from the mere absence of agency found in passive systems. Consequently, our framework posits that pathological agents operate within a valid, albeit unsuccessful, self-referential loop, suggesting that agency persists even in the state of regulatory failure.
%In conclusion, the differentiation between $V_{t'}^{S_{ant}}$ and $V_{t'}^{S}$ constitutes the functional interface where the system's self-referential loop actively attempts to regulate its own organization, a process inherently subject to both success and failure. While an external observer might view persistent maladaptation as a mere absence of control, the Internal Gap perspective reveals it as the activity of intrinsic dynamics $F_{S}$ attempting regulation that has failed. This distinction is critical; unlike a passive object determined exclusively by constituent micro-dynamics, pathological agents maintain a teleological orientation through its intrinsic constraints. By acknowledging that agency persists even during regulatory failure, this framework redefines dysfunction not as the negation of agency, but as the operation of a valid, albeit unsuccessful, self-referential loop.

% While an external observer might perceive persistent environmental maladaptation as a mere absence of function or control, the perspective of the Internal Gap allows us to interpret such states as the manifestation of active intrinsic dynamics $F_{S}$ exercising a regulatory capacity that has structurally or dynamically failed.

\subsection{The External Gap as the Operational Irreducibility of Intrinsic Dynamics
} \label{sec: Gap between internal and as-if}
%The External Gap as the Distinction between Description and Generation

While the system carries out its intrinsic regulation through $F_{S}$, the observer constructs an ``as-if'' model $F_{O}$ to describe this behavior in a coarse-grained form \cite{dennett1989intentional, mcgregor2017bayesian}.  Existing approaches typically frame behavior as the optimization of a specific objective function, such as free energy \cite{friston2010free} \footnote{As mentioned in the Introduction, the FEP can be interpreted both as the system's intrinsic dynamics and as the observer's ``as-if'' model \cite{ramstead2020tale}}, empowerment \cite{salge2014empowerment}, novelty or curiosity driven search\cite{schmidhuber1991possibility, oudeyer2007intrinsic, schmidhuber2010formal, burda2018large, lehman2011abandoning, mouret2011novelty, mouret2015illuminating, burda2018exploration}. Even approaches that attempt to internalize rewards, such as Homeostatic Reinforcement Learning \cite{keramati2011reinforcement, hulme2019neurocomputational, yoshida2017homeostatic, yoshida2023homeostatic} or Meta-Reinforcement Learning \cite{wang2016learning, wang2018prefrontal, horibe2024emergence, vafa2024evaluating}, replace external reward signals with internal variables but still operate as observer-defined descriptions of stability maximization. These frameworks successfully capture the system's behavior ``as if'' it were optimizing a well-defined goal under the observer's coarse-graining. 

Clarifying the interpretive gap therefore requires distinguishing these externally constructed descriptions from the system's own downward-causal loops. In biological systems, the External Gap typically persists as the system actively maintains its organization. However, circumstances exist where the observer's model $F_{O}$ and the system's intrinsic dynamics $F_{S}$ appear to converge. For example, in \textit{E. coli} chemotaxis, the intrinsic biochemical network $F_{S}$ is mathematically equivalent to an information-theoretically optimal ``as-if'' model $F_{O}$ for gradient navigation \cite{nakamura2022optimal}. Yet, such alignment does not by itself account for the generative process. A static match cannot reveal whether the system is actively maintaining its organization or simply following a fixed trajectory described by the observer. For this reason, evaluating an ``as-if'' model requires examining how the system modifies its intrinsic dynamics when environmental conditions change.

This adaptive plasticity is observed in empirical cases demonstrating that intrinsic dynamics $F_{S}$ are generative products of evolutionary or learning histories, capable of reorganization that a fixed ``as-if'' model cannot predict. For example, \textit{E. coli} evolution shows that correlations learned from environmental statistics (e.g., anticipating oxygen depletion from temperature rise) are rapidly reconfigured when experimentally inverted \cite{tagkopoulos2008predictive}. A deeper form of plasticity appears in the Dark-fly lineage: after 1,400 generations in darkness, \textit{Drosophila melanogaster} exhibits regression of the visual system and loss-of-function mutations in light receptor genes \cite{izutsu2012genome}. Similarly, artificial agents evolved for competitive foraging tasks continue to express matching behavior even when tested in non-competitive forced-choice environments \cite{seth2007ecology}, indicating that their intrinsic dynamics reflect historically embedded constraints rather than the task-specific objective imposed by the observer. Collectively, these biological and artificial examples illustrate that $F_{S}$ actively reshapes its parameters and structure based on historical viability, diverging from observer's ``as-if'' descriptions.

Importantly, the External Gap should not be conflated with mere predictive divergence. Even in the theoretical limit of descriptive convergence ($F_{O} \equiv F_{S}$), the significance of this alignment depends on the underlying generative process, presenting three distinct possibilities. In cases of structural self-organization typical of Deep Learning systems, the convergence arises because the observer has explicitly engineered $F_{S}$ to minimize divergence from a prescriptive $F_{O}$. Conversely, the observer may construct an ``as-if'' model $F_{O}$ that statistically reproduces the system's behavior without capturing the underlying intrinsic dynamics; here, the alignment is circumstantial, as environmental perturbations typically reveal the gap by dissociating the fixed description from the system's adaptive reorganization. Finally, an observer could theoretically construct an $F_{O}$ that fully incorporates the generative mechanism of the system's intrinsic dynamics $F_{S}$, although achieving such a comprehensive description for biological systems remains an empirically elusive goal. 
Thus, the External Gap signifies an organizational distinction rather than a statistical discrepancy: agency is defined by the causal efficacy of intrinsic constraints to shape the future via downward causation, a property that strictly distinguishes teleological self-organization from externally designed structural convergence.

\subsection{Summary}

In summary, the duality framework characterizes agency through the integration of two distinct interfaces. The internal interface defines the Internal Gap as the divergence between the system's intrinsic anticipation ($V_{t'}^{S_{ant}}$) and the realized state ($V_{t'}^{S}$), which functions as the error signal driving self-regulation. The external interface examines the relationship between these intrinsic dynamics ($F_{S}$) and the observer's ``as-if'' model ($F_{O}$), utilizing the External Gap to differentiate teleological self-organization from structural self-organization. Taken together, this framework establishes that a system possesses agency if and only if it satisfies two generative conditions: the instantiation of a self-referential loop established through the reciprocal interplay of internal coarse-graining and downward causation (Section \ref{sec: system}), and the maintenance of teleological self-organization that is treated as organizationally distinct from externally imposed design. (Section \ref{sec: interface and observer}).

\section{A Spectrum of Agency: Distinguishing Structural and Teleological Self-Organization}
\label{sec: spectrum}

To show how the duality framework distinguishes different forms of goal-directed behavior, we apply it to the representative systems listed in Table \ref{tab:spectrum}. Each system exhibits distinct relationships between microscopic dynamics $X_t$, internal coarse-grained variables $V_t^S$, and the observer's macroscopic variables $V_t^O$. These relationships determine whether the system processes information, mechanically regulates itself, or exerts agency through a self-generated gap. A central distinction is, this classification relies on identifying the generative process of the intrinsic dynamics $F_{S}$, specifically on whether the constraints governing the system are externally imposed or autonomously constructed. This distinction allows us to differentiate between structural self-organization, understood as the emergence of patterns under external constraints, and teleological self-organization, understood as the autonomous generation of the constraints themselves.

We begin with systems that fail to integrate the necessary components for agency. First, a Rolling Stone descending a potential energy gradient possesses no internal mechanisms for coarse-graining or self-modeling. Its dynamics are determined exclusively by constituent micro-dynamics; while an observer can model this perfectly as an optimization process $F_{O}$, no corresponding intrinsic dynamics $F_{S}$ exists, rendering the internal–external gap 'N/A'. Next, the Variational Autoencoder (VAE) \cite{kingma2013auto} performing unsupervised representation learning, successfully executes internal coarse-graining by compressing high-dimensional data $X_t$ into a latent variable $V_t^S$. However, this is structural self-organization: it lacks a downward causation loop, and the training objective driving its optimization is not self-generated but fully prescribed by an external designer. In contrast, the Watt Governor mechanically regulating speed via feedback \cite{baltieri2020predictions}, clearly exhibits a downward causation loop, where the macroscopic arm angle physically constrains the microscopic steam-valve state. However, this control is achieved through direct physical coupling rather than information compression (no internal coarse-graining). In all three cases, the absence of a complete self-referential loop, thus no agency is present.

The framework's utility becomes clearer when examining systems that appear to possess intrinsic dynamics $F_{S}$. Braitenberg vehicles \cite{braitenberg1986vehicles} exhibiting life-like behavior via simple sensor-motor coupling often evoke strong interpretations of agency by the observer. However, this is purely bottom-up dynamics without internal representation. The perceived teleology is ``Illusory'', existing only in the observer's $F_{O}$, leaving the Internal–external gap as 'N/A'. Next, we consider systems with a self-referential loop, a crucial distinction emerges between designed and self-organized agents. Deep Learning Systems, such as World Model architectures \cite{ha2018worldmodels} and Large Language Models (LLMs), utilize internal simulations to optimize tasks. They perform internal coarse-graining (e.g., VAEs or Transformer layers compressing data) and exert downward causation (e.g., latent states driving actions). However, this remains a form of structural self-organization because the generative origin of their intrinsic dynamics $F_{S}$ is designed. The $F_{S}$ is explicitly engineered to minimize divergence from an externally specified objective $F_{O}$ such as a reward function or externally provided training data. This dependency is particularly clear in systems fine-tuned through Reinforcement Learning from Human Feedback \cite{christiano2017deep, ouyang2022training}, where the optimization procedure systematically adjusts the intrinsic dynamics $F_{S}$ so that its behavioral outputs remain consistent with the evaluations provided by an external observer. Accordingly, although these systems exhibit an internal organization, the generative process shaping $F_{S}$ is determined by externally imposed criteria rather than by requirements arising from the system's own operational dynamics. In this sense, they function as advanced regulators that lack a self-generated internal–external gap.

In contrast, a Bacterium maintaining metabolic organization via intrinsic self-regulation exhibits all necessary components for minimal biological agency: internal coarse-graining of biochemical signals and a downward-causation loop that organizes bacterial movement. As discussed in Section \ref{sec: Gap between internal and as-if}, however, the presence of these structures does not determine whether the intrinsic dynamics $F_{S}$ exceed an observer's description $F_{O}$. This distinction becomes detectable only under environmental perturbations that modify task contingencies or statistical regularities. In bacterial chemotaxis, for example, perturbing environmental conditions can reveal whether the generative process of $F_{S}$ reorganizes to preserve metabolic organization or whether observed behavior remains consistent with a fixed ``as-if'' objective such as gradient optimization \cite{nakamura2022optimal}. When metabolic organization is maintained under such perturbations through reorganizations of $F_{S}$ that are not recoverable from any fixed $F_{O}$, the divergence between the two indicates a persistent internal–external gap. Under this criterion, the bacterium meets the operational conditions for Minimal Agency, with intrinsic dynamics organized around sustaining its own metabolic organization rather than implementing an externally specified task description.

Finally, Humans occupy the maximal end of this agency spectrum. In addition to sustaining metabolic organization, they construct reflective self-narratives and long-term commitments that influence the organization of their intrinsic dynamics $F_{S}$. Under environmental or social perturbations, reorganizations of $F{S}$ often diverge from any fixed observer model $F_{O}$, resulting in the most persistent internal–external gap within the spectrum. The generative process of $F_{S}$ also extends to the social domain, where internally coarse-grained models of other agents contribute to the regulation of interaction. These features indicate that human relational and regulatory capacities are grounded in their own intrinsic dynamics rather than in externally imposed task descriptions.

Accordingly, this distinction based on the generative process of intrinsic dynamics extends prominent frameworks such as the Free Energy Principle (FEP) and Good Regulator Theory (GRT). By differentiating teleological self-organization which the system autonomously constructs its own constraints through internal coarse-graining and downward causation, from structural self-organization and passive dynamics, the framework addresses the ambiguity inherent in purely descriptive approaches. In this view, agency is characterized not by the mere minimization of divergence or the possession of an ``as-if'' model ($F_{O}$), but by the system's capacity to sustain a self-generated internal–external gap ($F_{S} \neq F_{O}$) that cannot be reduced to external design or passive mechanics.

\begin{table}[ht]
    \centering
    \small % for preprint
    \begin{tabularx}{\textwidth}{ 
        >{\hsize=1.3\hsize\raggedright\arraybackslash}X 
        >{\hsize=1.05\hsize\raggedright\arraybackslash}X 
        >{\hsize=1.05\hsize\raggedright\arraybackslash}X 
        >{\hsize=1.0\hsize\raggedright\arraybackslash}X 
        >{\hsize=0.6\hsize\raggedright\arraybackslash}X 
    }    \toprule
    \multirow{2}{*}{\textbf{System}}
      & \multicolumn{2}{c}{\textbf{Internal}}
      & \multicolumn{1}{c}{\textbf{Interface}}
      & \multirow{2}{*}{\textbf{Is agent?}} \\ 
    \cmidrule(lr){2-3} \cmidrule(lr){4-4}
      & \textbf{Coarse-Graining} & \textbf{Downward causation loop} 
      & \textbf{Internal–External gap} & \\
    \midrule

    \textbf{A rolling stone} descending a potential energy gradient
      & \textbf{No}: A stone has no macro variable 
      & \textbf{N/A}
      & \textbf{N/A}
      & \textbf{No} \\
    \addlinespace 
    \addlinespace   
    \textbf{A Variational Autoencoder} performing unsupervised representation learning
      & \textbf{Yes}: compressing data into a low-dimensional latent variable
      & \textbf{No}: Macro variable does not regulate the system
      & \textbf{N/A}
      & \textbf{No} \\
    \addlinespace
    \addlinespace
    \textbf{A Watt Governor} regulating speed via mechanical feedback
      & \textbf{No}: No information compression; direct physical coupling
      & \textbf{Yes}: Macroscopic arm angle constrains microscopic valve state
      & \textbf{N/A}
      & \textbf{No} \\
    \addlinespace
    \addlinespace 
    \textbf{Braitenberg vehicles} exhibiting life-like behavior via simple sensor-motor wiring
      & \textbf{No}: Direct sensor-motor coupling; no internal representation
      & \textbf{No}: Purely bottom-up dynamics; no macroscopic constraint
      & \textbf{N/A} (Intrinsic dynamics is illusory)
      & \textbf{No} \\
    \addlinespace
    \addlinespace

    \textbf{Deep Learning Systems (e.g., World models, LLMs)} optimizing external objectives
    & \textbf{Yes}: Compressing data into latent variables or parameters (structural self-organization)
    & \textbf{Yes}: Internal states or context windows constrain future predictions and actions
    & \textbf{No}: $F_{S}$ is explicitly optimized to minimize divergence from external training data $F_{O}$
    & \textbf{No} \\
    \addlinespace
    \addlinespace  
    
     \textbf{Bacterium} maintaining metabolic organization via intrinsic self-regulation
      & \textbf{Yes}: Biochemical networks compress micro-states into signals (teleological self-organization)
      & \textbf{Yes}: Macroscopic signals (e.g. CheY-P level) constrain flagellar motion
      & \textbf{Yes}: Internal complexity $F_{S}$ is not always fully captured by specific observer models $F_{O}$
      & \textbf{Yes (Minimal)}: Reflective gap present but minimal \\
    \addlinespace
    \addlinespace

     \textbf{Humans} exercising reflective volition and social coupling
      & \textbf{Yes}: Generating complex self-narratives and long-term goals
    & \textbf{Yes}: Conscious intent constrains biological and motor dynamics
    & \textbf{Yes}: Subjective experience $F_{S}$ is not fully reducible to observation $F_{O}$
  & \textbf{Yes (Maximal)} \\
\bottomrule
    \bottomrule
    \end{tabularx}
    \caption{\textbf{A Spectrum of Agency: Classification of Systems under the Duality Framework.}
    This table summarizes how representative systems score on the key criteria of the duality framework: the presence of internal coarse-graining, the existence of a downward-causation loop, and whether an internal–external gap is maintained. These structural features are listed to show how different kinds of systems can be positioned along a continuous spectrum of agency.}
\label{tab:spectrum}
\end{table}

\section{Discussion} \label{sec: discussion}

\subsection{The Generative Turn in Understanding Agency}

This study reframes agency not merely as a property of a system nor solely as an attribution by an observer, but as a phenomenon emerging at the interface of generative intrinsic dynamics and epistemic interpretation. Central to this reframe is our grounding of agency in the generative process of a self-referential loop (Section \ref{sec: system}), which demonstrates that agency necessitates more than just minimizing a predictive divergence (as in FEP) or acting as a good regulator (as in GRT); it requires the autonomous construction of constraints through the reciprocal interplay of internal coarse-graining and downward causation. This generative perspective enables the distinction between the true teleological self-organization of biological organisms and the structural self-organization of current AI systems, a distinction made operationally visible through the Internal and External Gaps (Section \ref{sec: interface and observer})—metrics that define agency by the system's capacity to sustain a divergence between its intrinsic becoming and external description. As a result, by establishing this duality framework, we resolve the persistent ambiguity in defining agency that has historically conflated the ``as-if'' description of optimization with a system's own teleological organization.

However, this definition poses a significant challenge to current engineering paradigms. While artificial systems like World Models and LLM Agents incorporate mechanisms akin to internal coarse-graining (e.g., latent variables) and downward causation (e.g., context-driven generation), they remain fundamentally structural. Their intrinsic dynamics ($F_{S}$) are optimized to minimize the external gap ($F_{O}$), effectively collapsing the very interface that defines agency. This suggests that the path to creating artificial agents with genuine autonomy may not lie in larger models or better training data, but in a fundamental shift in optimization objectives: from minimizing external error to satisfying internally generated viability constraints. Future research should explore computational substrates where the maintenance of internal organization is a prerequisite for persistence, potentially drawing inspiration from open-ended evolution and autopoietic simulations.

A limitation of this framework is its current reliance on information-theoretic metrics which, while powerful, remain external evaluations (as discussed in Section \ref{sec: system}). Bridging the gap between these quantitative measures and the qualitative, phenomenological aspects of agency remains an open challenge. Furthermore, while we have outlined the generative conditions for agency, the specific implementation details for artificial systems that can autonomously acquire such teleological organization require further exploration.

\subsection{Future Direction: Social Entanglement via the Generative Gap}

\begin{figure}
    \centering
    \includegraphics[width=1.0\linewidth]{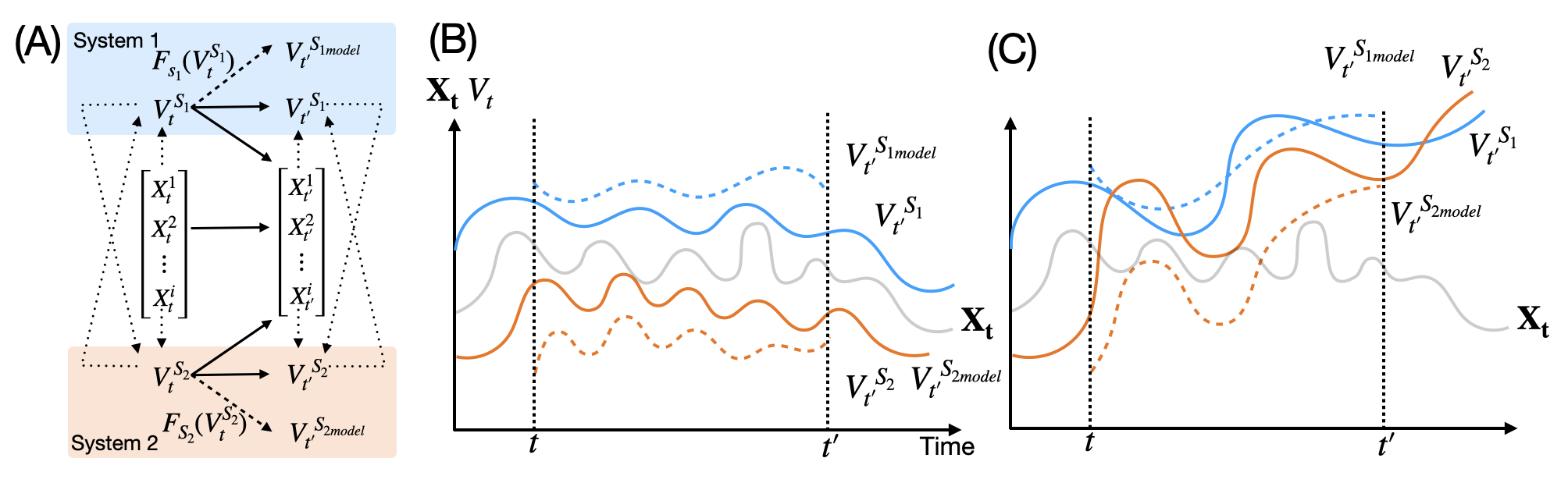}
    \caption{\textbf{Social Interaction via the Generative Gap.} 
    (A) The reciprocal modeling loop between two agents. (B) The regime of predictive convergence. (C) The regime of emergent entanglement forms a collective macro-variable. System 1 is shown in blue; System 2 in orange. General notations for variables ($X, V$) and arrow types follow the conventions defined in Figure \ref{fig: time evolution}.}
    \label{fig: agency between agents}
\end{figure}

% From the perspective of the proposed duality framework, the distinction between internal generative dynamics ($F_S$) and external coarse-grained observation ($F_O$) may offer a perspective on social interaction. In this context, the gap between $F_S$ and $F_O$ could be interpreted as a medium for social coupling rather than solely as a boundary of the individual. As shown in Figure \ref{fig: agency between agents}(A), this framework models social cognition as a reciprocal process. Here, each agent maintains its own teleological loop—characterized by downward causation from macrostate $V_t^{S_1}$ to microstates $X_t^i$—while simultaneously engaging in an ``as-if'' modeling process ($V_{t'}^{S_{2model}}$) to estimate the internal macrostate of the other. This formulation provides a schematic description that aligns with the concept of participatory sense-making \cite{de2007participatory}, proposing that social coordination might be analyzed through the coupled interaction of these internal and external loops.

From the perspective of the proposed duality framework, the distinction between internal generative dynamics ($F_{S}$) and external coarse-grained observation ($F_{O}$) offers an analytical perspective for social interaction. In this context, the gap between $F_{S}$ and $F_{O}$ functions as a medium for social coupling rather than solely as a boundary of the individual. As shown in Figure \ref{fig: agency between agents}(A), this framework models social cognition as a reciprocal process. Each agent maintains its own teleological loop, in which downward causation links the macrostate $V_t^{S_1}$ to the microscopic variables $X_t^i$, and at the same time engages in an ``as-if'' modeling process ($V_{t'}^{S_{2model}}$) to estimate the other's internal macrostate. This formulation not only aligns with the concept of participatory sense-making \cite{de2007participatory} but also offers a mechanistic account of how shared meaning can arise: it is shaped through the ongoing adjustment between an agent's intrinsic dynamics and its model of the other, with the resulting mismatch functioning as a regulative interface through which coordinated behavior is organized.

% This framework extends beyond descriptive metrics such as synchronization \cite{condon1974neonate,tognoli2007phi} and shared information \cite{trendafilov2020tilting, jahng2017neural} to classify the recursive estimation process (Figure \ref{fig: agency between agents}(A)) into two distinct dynamic regimes. The first is ``predictive convergence'' (Figure \ref{fig: agency between agents}(B)), where mutual modeling successfully tracks internal states while preserving individual intrinsic dynamics as the primary driver. The second is a structurally distinct state (Figure \ref{fig: agency between agents}(C)) where coupled dynamics form an autonomous macroscopic variable effectively decoupled from individual microscopic variables ($X_t$), a structure that may correspond to the collective organization observed in minimal interaction paradigms \cite{auvray2009perceptual, noy2011mirror, dumas2014human}.

Building on this account of how coordinated behavior is organized, the generative perspective further clarifies the dynamical structure of social interaction. It advances beyond descriptive metrics such as synchronization \cite{condon1974neonate,tognoli2007phi} and shared information \cite{trendafilov2020tilting, jahng2017neural}, which demonstrate that coupling occurs but not how a shared causal context is constituted. By treating social interaction as a reciprocal coupling between agentive loops—in which each agent maintains an ``as-if'' model of the other ($F_{O \to S}$)—the framework identifies distinct dynamical regimes that characterize the interface of agency. The first is ``predictive convergence'' (Figure \ref{fig: agency between agents}(B)), a regime of mutual predictability in which internal states are successfully inferred while each system retains its intrinsic dynamics. The second is ``emergent entanglement'' (Figure \ref{fig: agency between agents}(C)), a structurally distinct regime in which the coupled dynamics give rise to an autonomous macroscopic variable that is functionally decoupled from individual microscopic states ($X_t$). This formal characterization offers a unified account of collective patterns reported in paradigms such as the Perceptual Crossing Experiment \cite{auvray2009perceptual}, the Mirror Game \cite{noy2011mirror}, and the Human Dynamic Clamp \cite{dumas2014human}, and it suggests conditions under which the interaction itself may acquire viability constraints that are not reducible to either individual alone.

%Applying this framework to artificial systems points to a structural distinction in human-machine interaction. Humans may engage in reciprocal modeling by attributing an intrinsic dynamics ($F_S$) to an interlocutor exhibiting complex behavior \cite{seth2024conscious}. However, as discussed in Section \ref{sec: spectrum}, Large Language Models (LLMs) remain systems engineered to minimize external prediction error ($F_O \to F_S$), even though their behavioral complexity is now sufficient to pass the classical Turing test \cite{turing1950computing,jones2025large}. The present framework suggests that without the generative mechanics to sustain autonomous internal macroscopic goals ($F_S$), such interactions may differ structurally from the coupling between two teleological agents. This highlights a theoretical distinction between interaction with a system possessing self-generated intrinsic dynamics and interaction with a system optimizing for external convergence.

Applying this framework to artificial systems points to a structural distinction in human-machine interaction. Humans may engage in reciprocal modeling by attributing an intrinsic dynamics ($F_{S}$) to an interlocutor that exhibits complex behavior. However, attribution of such dynamics to systems lacking intrinsic generative mechanics constitutes a form of anthropomorphism that obscures their actual mode of operation\cite{seth2024conscious}. As discussed in Section \ref{sec: spectrum}, Large Language Models (LLMs) remain systems engineered to minimize external prediction error ($F_{O} \to F_{S}$), even though their behavioral complexity is now sufficient to pass the classical Turing test \cite{turing1950computing,jones2025large}. This suggests that interactions with current AI differ fundamentally from the coupling between two teleological agents: lacking self-generated constraints, AI cannot participate in the generative negotiation of meaning but only mirror human expectations. Recognizing this distinction is vital for designing ``Hybrid Agents'' that can genuinely collaborate rather than merely comply, potentially requiring architectures that maintain their own internal norms to foster productive friction and deeper social integration.

%Finally, the potential emergence of the autonomous macroscopic variable shown in Figure \ref{fig: agency between agents}(C) shares conceptual parallels with game-theoretic traditions \cite{von1944theory} and social niche construction \cite{yamagishi2016social}, where the constraints of the interaction emerge from the process itself. Future work could attempt to validate this transition from individual to social dynamics by quantifying these regimes using metrics of causal emergence, such as effective information \cite{hoel2013quantifying, rosas2020reconciling, rosas2024software}. This would involve testing whether the principles of internal coarse-graining and downward causation can be applied to describe collective forms of agency.

Finally, the potential emergence of the autonomous macroscopic variable shown in Figure \ref{fig: agency between agents}(C) connects our framework to broader questions of social evolution, offering a mechanistic bridge to game-theoretic traditions \cite{von1944theory} and social niche construction \cite{yamagishi2016social}. While classical game theory generally assumes pre-defined rules and static payoffs, our framework illustrates how "the game" itself emerges from interaction: the agents' coupled activities construct a shared context—manifested as the autonomous macroscopic variable—that subsequently constrains their behavior. This process is analogous to social niche construction, where constraints are not given but are products of the interaction itself. Future work could attempt to validate this transition from individual to social dynamics by quantifying these regimes using metrics of causal emergence \cite{hoel2013quantifying, rosas2020reconciling, rosas2024software}. By treating social structures as emergent macroscopic variables maintained by downward causation, this framework offers a scalable path to understanding how collective forms of agency arise from, and subsequently constrain, the interplay of individual hybrid agents.

\section{Conclusion}
The duality framework advances the study of agency by resolving the critical ambiguity between an observer's descriptive model and a system's intrinsic dynamics, a distinction often obscured in existing information-theoretic and dynamical approaches. By grounding agency in the generative process of a self-referential loop established through the reciprocal interplay of internal coarse-graining and downward causation, this framework provides a method to differentiate teleological self-organization from mere structural self-organization. This differentiation allows for a precise classification of systems along a spectrum of agency, encompassing even pathological states where regulation fails but teleological orientation persists, effectively separating biological organisms that autonomously construct constraints for self-maintenance from artificial systems whose dynamics are engineered to converge with external objectives. Ultimately, by defining agency through the persistence of the gap between intrinsic dynamics and external interpretation, the framework offers a scalable theoretical perspective that extends from individual biological autonomy to the emergent coordination observed in social participatory sense-making.

\vskip6pt

\enlargethispage{20pt}

%\ack{The authors thank Manuel Baltieri for insightful discussions that influenced this work. This work was supported by Special Postdoctoral Researcher Program in RIKEN Project code 202501094085 and JSPS KAKENHI Grant Numbers 24K20859 for K.H. and by JST, CREST Grant Number JPMJCR21P4 and JSPS KAKENHI Grant Numbers 24H01534 for K.S.}

%\ack{The authors would like to thank in particular Manuel Baltieri for feedback on a previous version of the manuscript. K.H. and K.S. contributed equally to this work. Both authors conceptualized the study, conducted the investigation and validation, created the visualizations, and acquired funding. K.H. prepared the original draft, while both authors reviewed and edited the final manuscript. This work was supported by the Special Postdoctoral Researcher Program in RIKEN Project code 202501094085 and JSPS KAKENHI Grant Numbers 24K20859 for K.H. and by JST, CREST Grant Number JPMJCR21P4 and JSPS KAKENHI Grant Numbers 24H01534 for K.S.}

% for preprint
\section*{Acknowledgments}
The authors would like to thank in particular Manuel Baltieri for feedback on a previous version of the manuscript. K.H. and K.S. contributed equally to this work. Both authors conceptualized the study, conducted the investigation and validation, created the visualizations, and acquired funding. K.H. prepared the original draft, while both authors reviewed and edited the final manuscript. This work was supported by the Special Postdoctoral Researcher Program in RIKEN Project code 202501094085 and JSPS KAKENHI Grant Numbers 24K20859 for K.H. and by JST, CREST Grant Number JPMJCR21P4 and JSPS KAKENHI Grant Numbers 24H01534 for K.S.

%%%%%%%%%% Insert bibliography here %%%%%%%%%%%%%%

\bibliographystyle{RS}
\bibliography{sample}
\end{document}